\documentstyle[sprocl]{article}

\input{psfig}

\newcommand{\txl}{T$\chi$L~}
\newcommand{\tr}{\mbox{tr}}
\def\Journal#1#2#3#4{{#1} {\bf #2}, #3 (#4)}

\def\ea{$\mbox{\it et al }$}

\def\NPB{{\em Nucl. Phys.} B}
\def\NPA{{\em Nucl. Phys. Proc. Suppl.} A}
\def\NPP{{\em Nucl. Phys. Proc. Suppl.} B}
\def\PLB{{\em Phys. Lett.}  B}
\def\PRL{\em Phys. Rev. Lett.}
\def\PRD{{\em Phys. Rev.} D}

\def\be{\begin{equation}}
\def\ee{\end{equation}}
\def\bea{\begin{eqnarray}}
\def\eea{\end{eqnarray}}

\begin{document}
{\hfill
\begin{minipage}{0pt}\scriptsize \begin{tabbing}
\hspace*{\fill} WUP-TH 00-14\\ 
\hspace*{\fill} HLRZ2000-12 \end{tabbing} 
\end{minipage}\\[8pt] 
}
\title{
PROBING THE  QCD VACUUM \\WITH FLAVOUR SINGLET OBJECTS:\\[.2cm]
$\eta'$ ON THE LATTICE
}

\author{
  T. STRUCKMANN$^a$, K. SCHILLING$^{a,b}$,  P. \"UBERHOLZ$^b$  \\
  N. EICKER$^b$, S. G\"USKEN$^b$, TH. LIPPERT$^b$, H. NEFF$^a$, B. ORTH$^b$,\\
  J. VIEHOFF$^a$, }

\address{
$^a$NIC, Forschungszentrum J\"ulich, 52425 J\"ulich and\\
DESY, 22603 Hamburg, Germany\\
k.schilling@fz-juelich.de}
\address{$^b$Fachbereich Physik, Bergische Universit\"at, Gesamthochschule
Wuppertal\\Gau\ss{}stra\ss{}e 20, 42097 Wuppertal, Germany\\[.4 cm]
(SESAM - T$\chi$L  Collaboration)}


\maketitle
\abstracts{ We present a study on the direct determination of the
  $\eta '$ mass on the full set of SESAM and \txl QCD vacuum configurations
  with 2 active flavours of Wilson fermions, at $\beta = 5.6$. We observe a
  definite dependency of the two-loop correlator on the topological charge
  sector.}

\section{Introduction}
Flavour symmetric hadronic states are of particular importance for the
understanding of nonperturbative aspects of quantum chromodynamics (QCD). In
fact they are expected to reveal important insight into the topological vacuum
structure induced by gluonic self interactions, as borne out in the famous
Witten-Veneziano formula~\cite{witten-veneziano} that relates the flavour
octet/singlet mass splitting, $M_0^2$~\footnote{Upper (lower) case letters
  refer to masses in physical (lattice) units.}, to the topological
susceptibility,
$\chi_q$, in the large-$N_c$ limit of the theory:
\begin{equation}
M_0^2 := M_{\eta'}^2 - M_8^2 = 2 N_f \chi_q/f_{\pi}^2 \; ,
\label{eq:witten}
\end{equation}
with $N_f$ being the number of active flavours and $f_{\pi}$ the pion
decay constant.

While glueballs have been the target of many {\it ab initio} lattice
investigations over the past fifteen years~\cite{glueballs} much less
attention has been paid to the direct calculation of flavour singlet states
{\it with fermionic content} such as the $\eta '$ meson.  The reason is
obvious: the extraction of flavour singlet mesons from the lattice is even
more challenging than the one of glueball states, due to the substantial cost
of computing quark  (rather than Wilson) loop correlators.  In fact such
Zweig-rule forbidden diagrams (in general misleadingly called `disconnected
diagrams') consist of two quark loops connected via gluon lines only and need
to be calculated in the momentum zero state, which amounts to the
computationally very expensive evaluation of the trace of the inverse Dirac
operator.  The two pioneering studies in the field used quenched vacuum
configurations,  with Wilson fermions and wall sources~\cite{kuramashi} and
staggered fermions with stochastic sources~\cite{kilcup}. 
\begin{table}[t]
\caption{Lattice parameters and numbers of independent vacuum field configurations and
stochastic sources,
$N_e^{ll}$ . Last two columns: t-ranges for the single exponential
fits to the octet and singlet channel correlators.}
\begin{center}
\begin{tabular}{|ccccccc|}
\hline
$\kappa_{sea}$ & $m_{\pi}/m_{\rho}$ & $L^3*T$ & $N_{e}^{ll}$ 
& $N_{conf}$ &$G_8$-Fit & $G_{\eta'}$-Fit\\ 
\hline
0.1560 & 0.834(3)  & $16^3*32$ & $400$ & $195$ & $12-16$ & $6-9$\\
\hline
0.1565 & 0.813(9) & $16^3*32$ & $400$ & $195$ & $13-16$ & $6-9$\\
\hline
0.1570 & 0.763(6) & $16^3*32$ & $400$ & $195$  & $12-15$ & $6-9$\\
\hline
0.1575 & 0.692(10) & $16^3*32$ & $400$ & $195$  & $12-15$& $6-9$\\
\hline
0.1575 & 0.704(5)  & $24^3*40$ & $400$ & $156$  & $12-15$& $6-9$\\
\hline
0.1580 & 0.574(13) & $24^3*40$ & $100$ & $156$  & $12-15$& $6-9$\\
\hline
\end{tabular}
\end{center}
\label{tab:simdat}
\end{table}

This situation will change with the imminent advent of near-Teraflops
computers that promise both to provide (a) sufficient sampling rates for full
QCD vacuum configurations and (b) the analysing power to deal with the above
mentioned loop-loop correlators.  Albeit present day simulations are still
restricted to the case of two active flavours, $N_f = 2$, it is of great
interest to fathom state-of-the art techniques to deal with Zweig rule
forbidden diagrams. First results on flavour singlet masses  were presented recently
by CP-PACS~\cite{AliKhan:1999zi} and UKQCD~\cite{Michael:1999rs}.

In the present paper we present a study on the $\eta'$ mass using the full set
of QCD configurations generated by SESAM ($16^3\times 32$, `small
lattice') and T$\chi$L~\cite{sesam_txl} ($24^3\times 40$, `large lattice'), both at
$\beta = 5.6$ with standard Wilson action (see Table~\ref{tab:simdat}).
\section{Disconnected diagrams }
\subsection{The problem}

\label{section:observables}
We consider the pseudoscalar flavour singlet operator in a flavour symmetric
theory
\begin{equation}
\eta'(x) =  \sum_{i=1}^{N_f} \overline{q}_i(x) \gamma_5 q_i(x) \ ,
\label{eq:operaror}
\end{equation}
with $N_f$ flavours. By the usual  Wick contraction it leads
 to the flavour singlet propagator in terms of the inverse Dirac operator,
$M^{-1}$:
\begin{eqnarray}
\label{eq:greens}
 C_{\eta'}(0|x) & \sim & N_f \tr((M^{-1}(0|x))^{\dagger}M^{-1}(0|x)) \nonumber \\
 && - N_f^2 \tr(\gamma_5M^{-1}(0|0))^{\dagger} \tr(\gamma_5M^{-1}(x|x)) \; ,
\end{eqnarray}
which is a sum of fermionic connected and disconnected terms with traces to be
taken in the spin and colour spaces. In the rest of the paper we shall refer
to them as `one-loop' and `two-loop' contributions, respectively.
The momentum zero projection 
\begin{equation}
C_{\eta'}(t) \equiv \langle \eta'(t)\eta'(0)\rangle_{conn}
     - \langle \eta'(t)\eta'(0)\rangle_{disc} 
\label{eq:gvont}
\end{equation}
is expected to decay exponentially, 
     $\sim \exp(-m_{\eta'}t)$ and reveal the flavour singlet mass, $m_{\eta'}$.
On an antiperiodic (of length $T$ in time) lattice one should encounter the usual
$\cosh$ behaviour at large values of $t$ and $T-t$
\begin{equation}
G(t) \rightarrow \exp(-m_{\eta'}t) + \exp(-m_{\eta'}(T-t))\; .
\end{equation}
From this parametrization local effective masses $m_{\eta'}(t)$ can be retrieved
by solving the implicit equations 
\begin{equation}
\frac{G(t+1)}{G(t)} = 
\frac{\exp(-m_{\eta'}(t+1))+\exp(-m_{\eta'}(T-t-1))}{\exp(-m_{\eta'}t) +
 \exp(-m_{\eta'}(T-t))}\; ,
\label{eq:effmasses}
\end{equation}
where the {\it l.h.s.} ratios are  taken from the lattice.
\subsection{Diagonal improved stochastic estimator}
\label{section:estimator}
Obviously the momentum zero projection embodied in eq.~\ref{eq:greens}
requires the knowledge of the traces, $\tr(\gamma_5M^{-1}(x|x))$, on each time
slice. But their exact evaluation is far beyond the scope of present
computers.

One therefore takes resort to  a stochastic estimate  of the
disconnected diagrams by introducing an ensemble of $N_e$ nonlocal sources,
$\phi(x)^a, a = 1, \dots N_e$ with uncorrelated stochastic entries
on the lattice sites $x \in \mbox{colour} \times \mbox{spin} \times R_4$ space
such that in the limit $N_e \rightarrow \infty$ the ensemble averages show the
following limiting behaviours
\begin{eqnarray}
\langle \phi(x) \rangle \equiv \frac{1}{N_{e}}\sum_{a=1}^{N_{e}}
\phi^a(x)  & \rightarrow &  0  \\
 \langle \phi(x)^{\dagger} \phi(y) 
\rangle  & \rightarrow  &  \delta_{x,y} \ . \label{eq:limit}
\end{eqnarray}
In our case, we chose $Z_2$ noise for the components of the source vectors.
By solving the Dirac equation
\begin{equation}
M(z,x)\xi(x)  = \phi(z)
\label{eq:dirac}
\end{equation}
on the stochastic sources $\phi(z)$ one retrieves a bias free
estimate for the inverse Dirac operator in the large $N_e$ limit
\begin{equation}
\label{eq:genetax}
\langle \phi(y)^{\dagger} \xi(x) \rangle  =   \sum_z M^{-1}(z,x) \langle 
\phi(y)^{\dagger} \phi(z) \rangle 
  \rightarrow   M^{-1}(y,x) \; ,
\end{equation}
and hence for the one-loop term, $\tr(\gamma_5M^{-1})$.
\begin{figure}[t]
\centering{\epsfig{figure=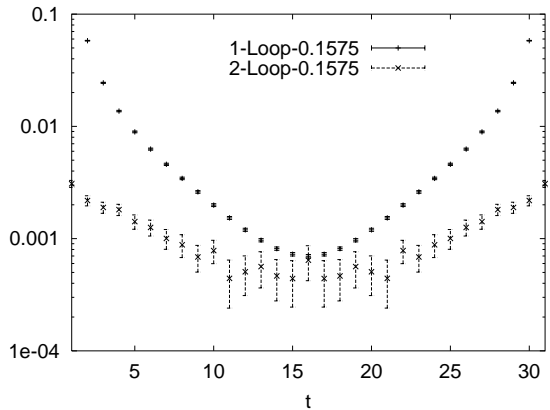,width=8cm,angle=270}} 
\caption{Correlation functions on the small lattice at the lightest sea quark
  mass, $\kappa = 0.1575$, with  local  sources.
Upper (lower) curves refer to one-loop (two-loop) contributions.}
\label{fig:correlators}
\end{figure}
But beware of the subleading terms when computing  components of $M^{-1}$
as we do here (actually  in spin space); for they might be  affected
 by the dominant diagonal ones in so far the latter are not
sufficiently rejected  by our approximation to the Kronecker $\delta$ in
eq.~\ref{eq:limit}. So we improve the estimate on nondiagonals,
$M^{-1}(y,x)$ ($x \ne y$), by subtracting out, from the expression on the {\it
  l.h.s.} of eq.~\ref{eq:genetax}, those leading error terms by redefining the
estimator to be~\cite{viehoff:stochastic}:\\
\begin{eqnarray}
\label{eq:impstoch}
\langle \phi(y)^{\dagger} \xi(x) \rangle - 
M^{-1}(x,x) \langle \phi(y)^{\dagger} \phi(x) \rangle
& = & M^{-1}(y,x)  \nonumber \\ & &+
  \sum_{z\neq x,y} M^{-1}(z,x) \langle \phi(y)^{\dagger} 
\phi(z) \rangle \; .
\end{eqnarray}

Note that in this improvement step, the subleading term on the {\it l.h.s.}
again is gained by SET. Throughout this work we shall apply
eq.~\ref{eq:impstoch} in order to achieve improved estimates to $\tr(\gamma_5
M^{-1})$. We mention in passing that an alternative route would be to work in
a spin explicite mode by using spin projected
sources~\cite{viehoff:stochastic}~\cite{wilcox}.
\begin{figure}[t]
\centering{\epsfig{figure=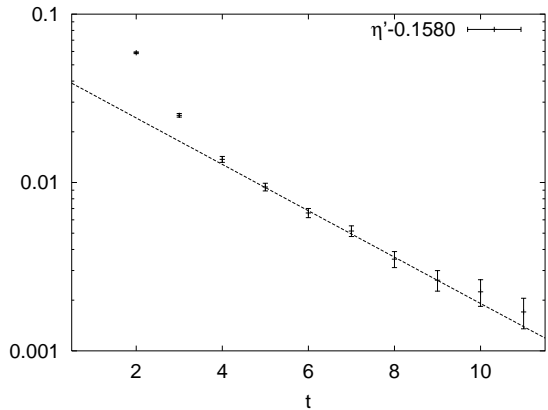,width=8cm,angle=270}}
\caption{$\eta'$ propagators ($\kappa = 0.158$) with local source and sink
on the large lattice.
\label{fig:tfit}}
\end{figure}
\subsection{Quality of the signal}
In Table~\ref{tab:simdat} we collected the characteristic parameters of our
present analysis.  We have used
five different sea quark masses and two different lattice sizes to have some
control on finite-size effects. While the number of vacuum configurations
varies from 156 to 195, the number of independent stochastic sources has been
chosen to be 400 (100) for the small (large) lattices.  The pseudoscalar and
vector mass ratios quoted refer to our final spectrum analysis~\cite{orth}.
The results presented in this talk are based on local sources only.  The errors
quoted are statistical and have been obtained by jackknifing.

As a result we obtain two-loop correlators, $C_{disc}(t) $, with
statistical quality illustrated for our lightest quark mass on the small
lattice in Fig.~\ref{fig:correlators}. The plot also contains the connected
correlator, $C_8(t)$, corresponding to the nonsinglet pseudoscalar meson, in
order to demonstrate the reduced accuracy of the (Zweig-rule forbidden)
two-loop contribution.  At the time separation $t \simeq 10$, we are faced
with a statistical error of order $\approx 25$ \% on the latter.

Given this situation it is practically impossible to establish an effective
mass plateau in the $\eta'$ channel, $C_{\eta'}(t)$, (eq.~\ref{eq:greens}).
For further analysis we made single exponential fits to $C_{\eta'}(t)$ and
$C_8(t)$ over the t-ranges listed in Table~\ref{tab:simdat}.  The quality of our
fits is illustrated for the lightest quark mass on the large lattice in
Fig.~\ref{fig:tfit}.
\begin{figure}[t]
\psfig{figure=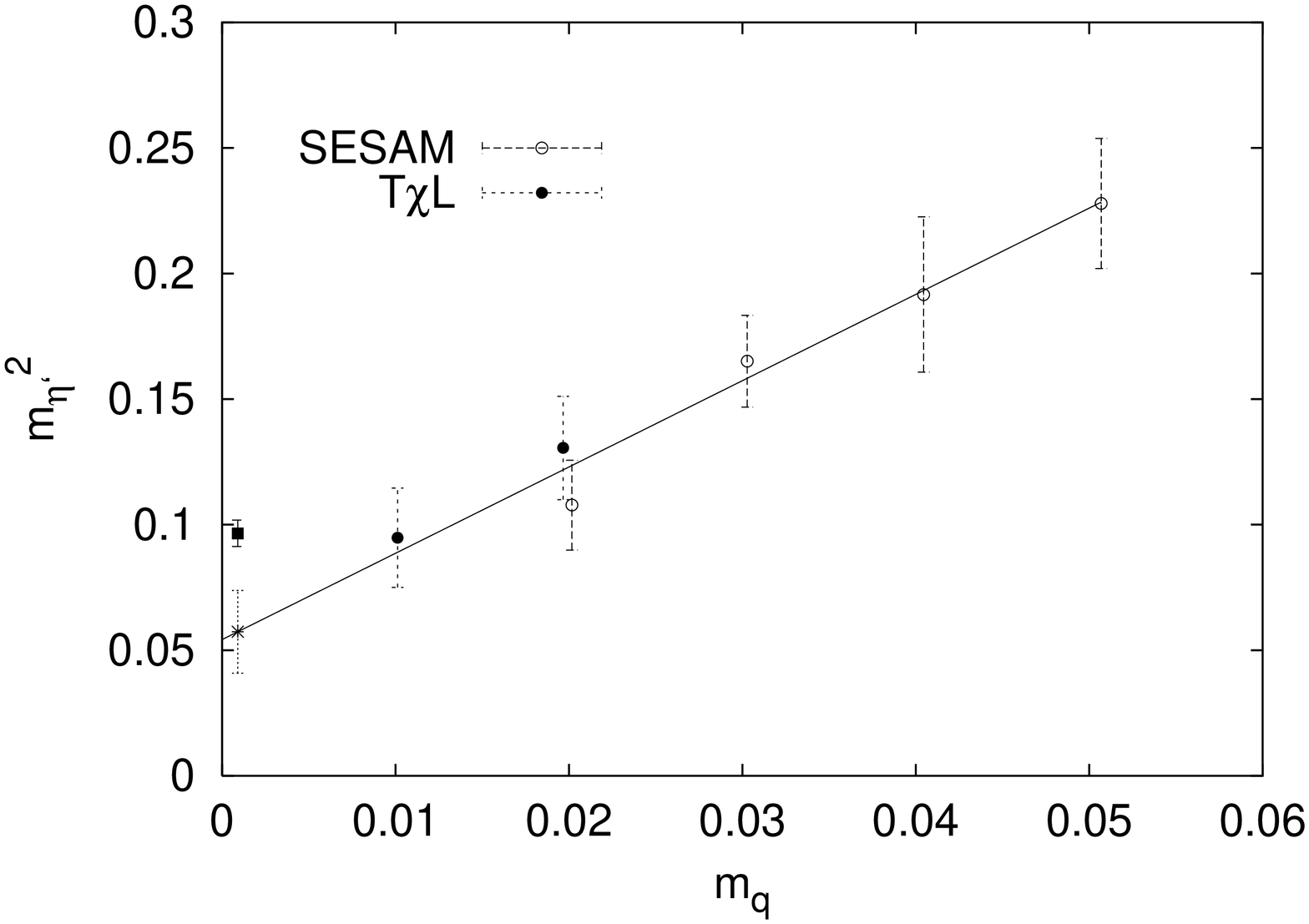,width=6cm,angle=270}
\vskip -4.2cm
\hfill
\psfig{figure=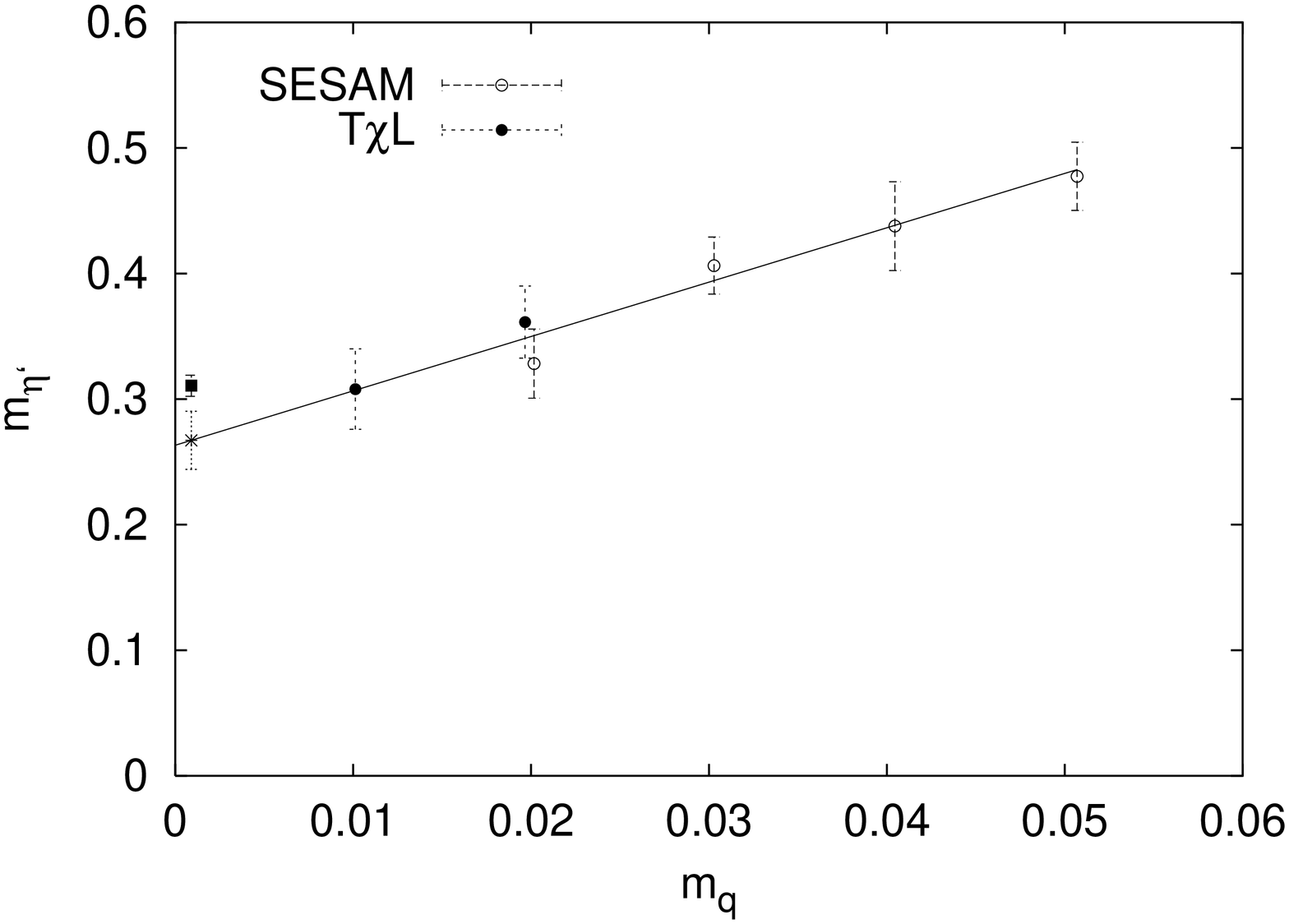,width=6cm,angle=270}
\caption{Chiral extrapolation of $m_{\eta'}^2$ (left) and 
  $m_{\eta'}$ (right), both linear in the quark mass.  Masses in lattice
  units. Full squares: pseudoexperimental value, according to
  eq.~\ref{eq:reduced}.}
\label{fig:xeta}
\end{figure}
\section{Physics results}
\subsection{Chiral extrapolations}
Because of the well-known technical limitations of the hybrid Monte Carlo
algorithm~\cite{1999:Kennedy}, the SESAM and T$\chi$L configurations
correspond to two mass degenerate light sea quark flavours ($N_f = 2$), with
the unrenormalized mass value (in lattice units)
\begin{equation}
m_q = 1/2(\kappa^{-1} - \kappa_{c}^{-1}) \;.
\end{equation}
From our previous light spectrum analysis~\cite{sesam:masses}, we quote the
lattice spacing 
\begin{equation}
a_{\rho}^{-1}(\kappa_l) = 2.302(64)\mbox{GeV} 
\end{equation}
and the critical and light quark $\kappa$ values:
\begin{equation}
\kappa_c = .158507(44)\quad, \quad \kappa_{light}= .158462(42) \; .
\end{equation}
From our data we cannot decide whether it is $m_{\eta'}^2$ or $m_{\eta'}$ that
follows a linear behaviour in the quark mass: the two fits shown in
Fig.~\ref{fig:xeta} work equally well, with $\chi^2/d.o.f. \simeq
{\cal O}(1)$. Note that we make {\it no distinction} between sea and valence
quarks, as we choose the quark masses in the fermion loops to equal  the sea
quark masses (symmetric extrapolation in the sense of
ref.~\cite{sesam:masses}).
\begin{figure}[h]
\begin{center}
 \epsfig{figure=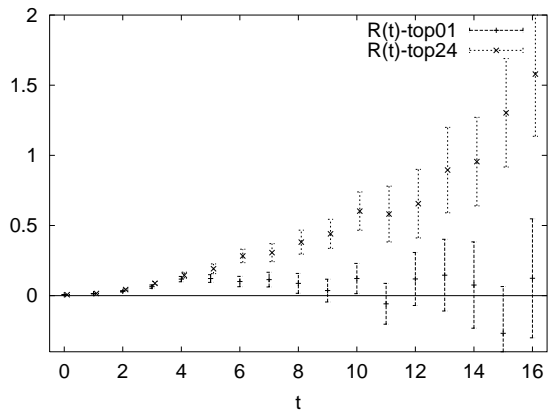,width=8cm,angle=270}
\caption{The sensitivity of the disconnected   diagram, 
$R_Q(t) := C_{disc}(t)/C_8(t)$, when subsampling the vacuum
  configurations according to topological charge, $|Q|$.}
\label{fig:topratio}
\end{center}
\end{figure}

In our $N_f = 2$ world, according to eq.~\ref{eq:greens}, we would not expect
to encounter the full effect of Zweig rule fordbidden diagrams, and hence we
anticipate to underestimate the real world $\eta '$ mass.
From the experimental mass splitting
\begin{equation}
M_{0}^2 = M_{\eta'}^2 - M_{8}^2 \; ,
\end{equation}
we therefore compute, in the spirit of the Witten-Veneziano formula
 eq.~\ref{eq:witten},
 the `pseudoexperimental' value in the $N_f = 2$ world:
\begin{equation}
M_{\eta'}^2(N_F=2) = 2/3M_0^2 + M_{\pi}^2 = (716 \mbox{MeV})^2 \; .
\label{eq:reduced}
\end{equation}
This value corresponds in lattice units to the full squares in the two
alternative chiral extrapolations shown in Fig.~\ref{fig:xeta}. 
To compare: our lattice analysis yields at the light quark mass
\begin{equation}
M_{\eta'}^2 = (551(85) \mbox{MeV})^2 \quad  \mbox{and} \quad
M_{\eta'} = 615(53) \mbox{MeV} \; ,
\end{equation}
by use of a  linear ansatz in $m_{\eta'}^2(m_q)$ and $m_{\eta'}(m_q)$,
respectively.
\subsection{Impact of topology}
We expect two-loop diagrams to be sensitive to the topology of the vacuum.  A
simple check is to look for a dependency of the ratio of disconnected and
connected correlators, $R_Q(t)$, on the size of the topological charge $|Q|$,
as determined in ref.~\cite{tunelling}.  In Fig.~\ref{fig:topratio}, we have
plotted, for $\kappa_{sea} = .1575$ and the small lattice, the data with cuts
according to $|Q| \leq 1.5$ (top01) and $|Q| > 1.5$ (top24).  We definitely
find a dependency of $R_Q$ on $|Q|$.  Note that the disconnected piece
vanishes in the vacuum sector with small values of $|Q|$!
\begin{figure}[t]
\begin{center}
  \epsfig{figure=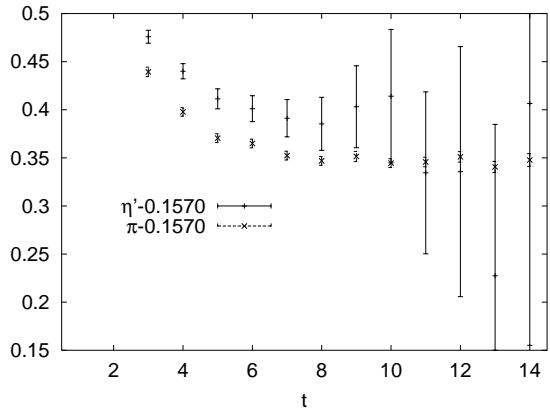,width=8cm,angle=270}
\caption{The effective local  octet ($\pi$) and singlet ($\eta'$)
  masses after Wuppertal type  smearing on source and sink.}
\label{fig:plateau}
\end{center}
\end{figure}
\section{Discussion and outlook}
The analysis presented here looks rather promising, but leaves room for
further investigations:

{\bf Smearing.}  The $\eta'$ correlator being a difference of one-loop and
two-loop terms it is of great importance to {\it attain early asymptotics} of
the flavour singlet correlator.  But with local sources the connected (flavour
octet) correlator does require analysis at $t$-values beyond $12$, in view of
appreciable excited state contributions.  Yet for the flavour singlet
situation, we had to analyse in the t-range 6 to 9, for lack of statistics.
Thus it appears very promising to improve the ground state overlap by use of
source and sink smearing; for this will help to establish  mass plateaus
in the flavour singlet channel and hence reduce errors.  Indeed, we have observed
mass plateaus appearing at $t \simeq 4$ to $5$, as illustrated in
Fig.~\ref{fig:plateau}. In a forthcoming paper, we shall elaborate on this
point~\cite{toappear}.

\begin{figure}[h]
\begin{center}
 \epsfig{figure=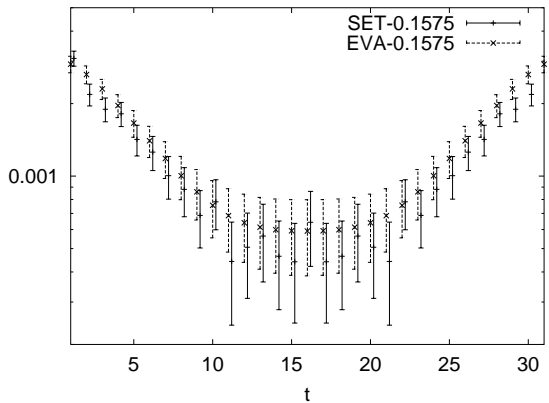,width=8cm,angle=270}
\caption{Saturation of the two-loop correlator with the 300
  lowest eigenmodes of the operator $(\gamma_5 M)$, at the smallest sea quark
  mass on the small lattice. Symbols: crosses from EVA, horizontal lines from
  SET.}
\label{fig:spectral}
\end{center}
\end{figure}

{\bf Spectral methods.} Another possible direction to go is to construct the
two-loop correlator from the low lying eigenfunctions of the Dirac operator.
For illustraton we show in Fig.~\ref{fig:spectral} the result of an attempt to
saturate the spectral representation of the two-loop correlator by the 300
lowest eigenmodes of $(\gamma_5 M)$~\cite{neff}. We find very nice agreement
between the results from this eigenmode approach (EVA) from SET. This is very
encouraging, since in the deeper chiral regime EVA  will become even 
more efficient whereas  the performance of SET  will deteriorate.
\section*{Acknowledgements} 
T.S., H.N., and B.O. thank the DFG-Graduiertenkolleg ``Feldtheoretische und
Numerische Methoden in der Statistischen und Elementarteilchenphysik'' for
support.  The HMC productions were run on an APE100 at NIC Zeuthen and INFN
Roma while the loop computations were carried out on APE100 systems at DESY
Zeuthen and at the University of Bielefeld.  We are grateful to our colleagues
F. Rapuano and G.  Martinelli for the fruitful T$\chi$L-collaboration.
Analysis was also performed on the CRAY T3E system of ZAM at Research Center
J\"ulich. K.S. thanks the organizers of Confinement2000, in particular
Prof. H.~Suganuma, for inviting him to
their very stimulating symposion.

\end{document}